\documentclass[a4paper,12pt,3p,twocolumn,times]{elsarticle}

\usepackage{amssymb}
\usepackage{amsmath}
\usepackage{booktabs}

\usepackage{hyperref}
\usepackage{cases}

\def\equationautorefname~#1\null{Eq. (#1)\null}

\journal{Nuclear Instruments and Methods in Physics Research Section A}

\begin{document}
	
	\begin{frontmatter}
		
		
		
		\title{New data acquisition system of the Lintott magnetic spectrometer at the S-DALINAC%\tnoteref{t1}
		}
		
		%\tnotetext[t1]{This document is the results of the research project funded by the Deutsche Forschungsgesellschaft (SFB 1245).}
		

		\author[]{M. Singer}
		\author[]{U. Bonnes}
		\author[]{A. D'Alessio}
		\author[]{M. Hilcker}
		\author[]{P. von Neumann-Cosel\corref{cor1}}
		\ead{vnc@physik.tu-darmstadt.de}
		\author[]{N. Pietralla}
		
		\cortext[cor1]{Corresponding author}
		
		\address%[]
		{Institut f\"ur Kernphysik, Technische Universit\"at Darmstadt, 64289 Darmstdt, Germany}
		
		\begin{abstract}
			A new data acquisition system for the high resolution magnetic spectrometer Lintott at the superconducting Darmstadt electron linear accelerator S-DALINAC was developed. 
			It allows inclusive and coincidence electron scattering experiments with event rates up to 10 kHz.
		\end{abstract}
		
		
		
		
	\end{frontmatter}

	
	\section{Introduction}
	
	The Institute for Nuclear Physics at the Technical University Darmstadt has a long-standing history in nuclear structure studies using electron scattering at low momentum transfers \cite{Pietralla2018} with a focus on the magnetic dipole response of nuclei \cite{Heyde2010}.
	A central research instrument has been the Lintott spectrometer \cite{Schull1977} with a beam line designed for dispersion matching \cite{Walcher1977}, thus allowing high-resolution measurements. 
	
	In 2006, the original focal plane detector consisting of overlapping scintillators was replaced by a system based on silicon strip detectors \cite{Lenhardt2006}, which considerably improved the data taking because the long and cumbersome measurements of the varying scintillator efficiencies due to radiation damage could be avoided.
	For maximum data rates no list mode data were recorded but the position information from the triggered strip was directly converted into an energy histogram.
	In recent years, a number of current nuclear structure questions have been successfully addressed with this system including the astrophysically important transitions to the Hoyle state state in $^{12}$C \cite{Chernykh2007,Chernykh2010} and to the first excited state in $^9$Be \cite{Burda2010}, investigations of mixed-symmetry states \cite{Burda2007,Walz2011,Scheikh2013,Scheickh2014}, of shape coexistence \cite{Kremer2016} or a precision measurement of the first excited $2^+$ state in $^{12}$C testing novel ab initio-based shell-model approaches \cite{DAlessio2020}. 
	
	However, the recent implementation of a high-energy scraper in the beam line \cite{Jurgensen2019}, which allows in principle extremely high-resolution measurements opens a new opportunity to increase the energy resolution even further by offline analysis requiring list mode data. This has led to the need of a redesign and further development of the data acquisition presented here.        
	
	\section{Lintott Spectrometer}
	
\begin{figure}
		\centering
		\includegraphics[width=\columnwidth]{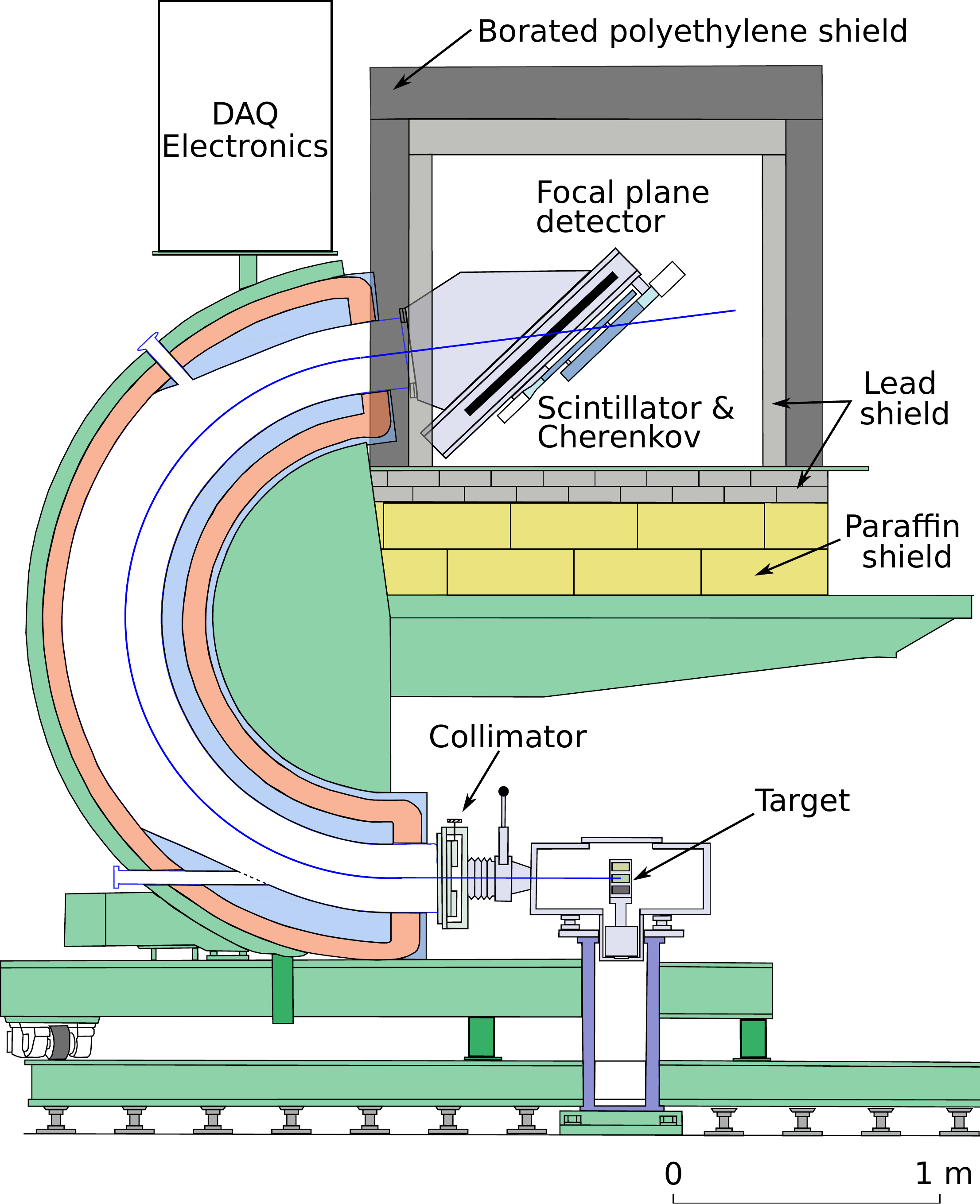}
		\caption{Cross section of the Lintott magnetic spectrometer.}
		\label{fig:spectrometer}
	\end{figure}

	The Lintott spectrometer is a vertically placed magnetic spectrometer built in the 1970s \cite{Schull1977}.
	A cross section is presented in \autoref{fig:spectrometer}.  
	The most important characteristics of the spectrometer are summarized in Tab.~\ref{tab:lintott}.
	A special operational mode of the Lintott spectrometer the so-called dispersion matching, where the dispersion of the extracted beam is matched to the dispersion of the spectrometer. Basically, the momentum spread of the electron beam is transformed into a spatial distribution on the target. The width of that distribution matches the different path lengths of the electrons inside the dipole magnet due to the momentum spread, such that they focus to a point at the focal plane where the detector is placed \cite{Walcher1977}. In this way the energy spread of the electron beam can be eliminated.
	
	Electrons scattered into the spectrometer port with a solid angle of 6 msr are collimated by vertical and horizontal slits to improve the energy resolution.%, reducing the effective solid angle to about 2.5 msr. 
	The dipole magnet separates the electrons by their momentum with a dispersion of 3.72 cm/\% in the focal plane. 
	The electron position in the focal plane is measured by a silicon strip detector located inside the vacuum system of the spectrometer.

	\begin{table}%[t]
		\caption{Spectrometer characteristics defined by design. The energy resolution is limited by the position resolution of the silicon strip detector. }
		\begin{tabular}{ll}
			\toprule
			Characteristic & Value \\	
			\midrule	
			Momentum acceptance & $\pm 2$\% \\
			Max. solid angle & $6$ msr \\
			Max. momentum & $120$ MeV/c \\
			Dispersion & $3.72$ cm/\% \\
			Energy resolution & $<1\cdot 10^{-4}$ (FWHM)\\
			Spectrometer angles & $69$$^{\circ}$-$165$$^{\circ}$, $12$$^{\circ}$ steps\\
			Max. event rate (DAQ) & 10 kHz \\	
			\bottomrule	
		\end{tabular}
		\label{tab:lintott}
	\end{table}

	After passing the position detector, the electrons leave the spectrometer through a mylar foil and hit the trigger scintillator and the Cherenkov detector. The latter serves for the suppression of the background due to gamma-rays mainly originating from the target.
	The read-out electronics for the detector system and the supply voltage for the photo-multipliers, the silicon strips and other electronics are placed on the top of the spectrometer shielded from radiation during the experiment. 

	\subsection{Silicon strip detector}
	
\begin{figure}[t]
		\centering
		\includegraphics[width=0.45\textwidth]{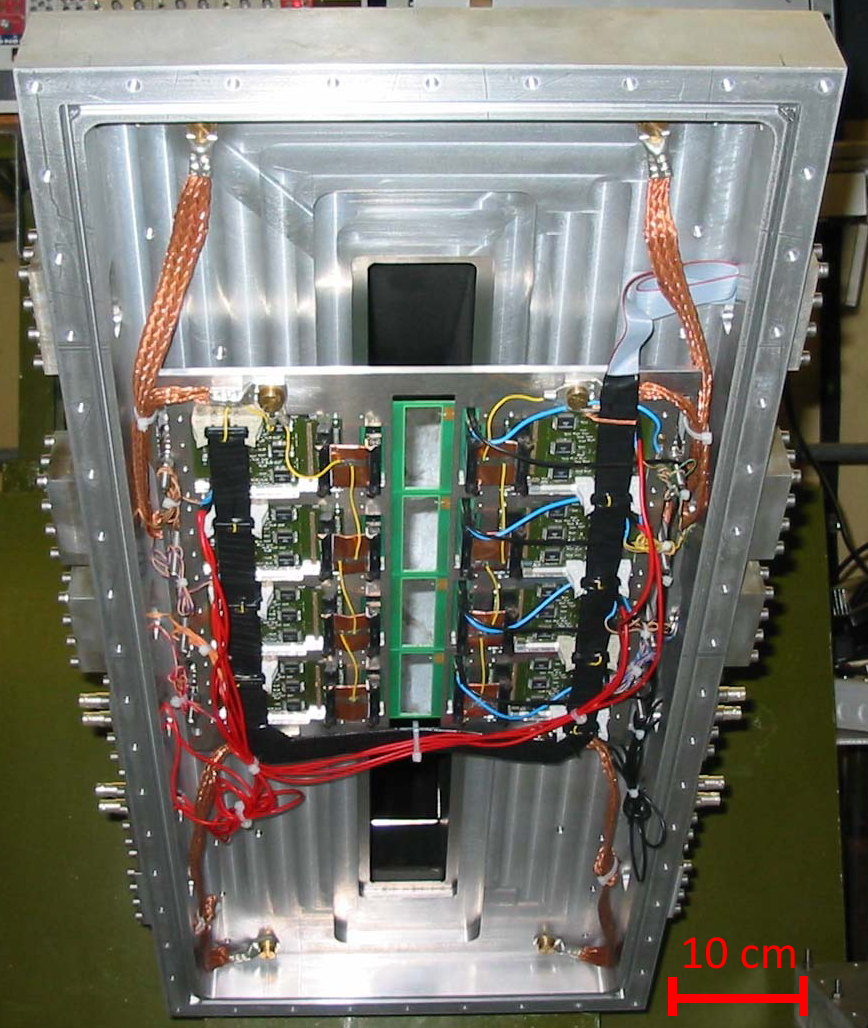}
		\caption{Focal plane detector system inside the vacuum chamber of the 169$^\circ$ spectrometer. 
			In the center 4 silicon strip detectors, each 96 channel wide, are visible. 
			Around them are Gassiplex chips on the PCB. 
			The 8 multiplexed output signals and the control signals for the Gassiplex chips are connected through vacuum feedthroughs to the readout electronics.}
		\label{fig:lintott_detektor}
	\end{figure}

\begin{figure}[t]
		\centering
		\includegraphics[width=0.47\textwidth]{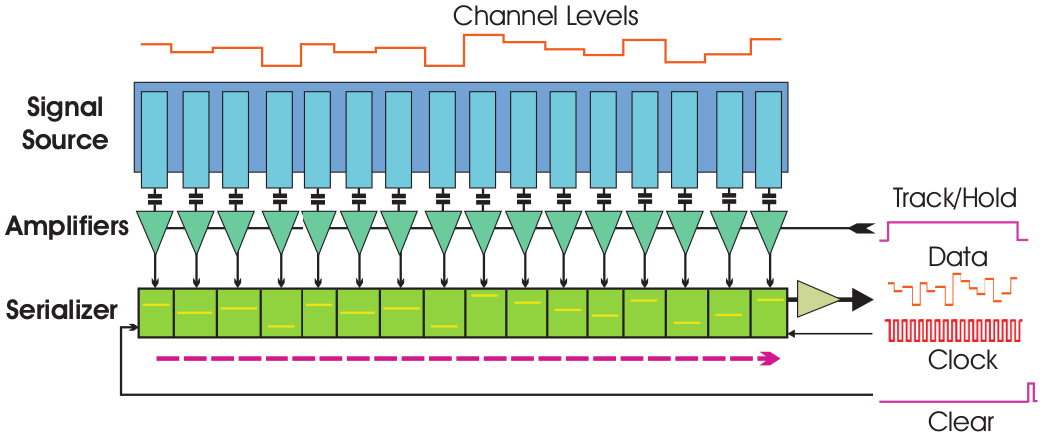}
		\caption{Read-out scheme of a Gassiplex chip. 
			A Hold signal let the Gassiplex chips keep the voltage levels depending on the collected charge inside the silicon strips. 
			At every new Clock pulse they are successively put to the output driver. 
			Finally, a Clear signal sets the electronics back to the live state 
			(adapted from Ref.~\cite{Lenhardt2006}).}
		\label{fig:gassiplex_readout}
	\end{figure}

	The focal plane detector consists of four silicon strip modules with 96 strips each. 
	One strip has a width of 650 $\mu$m and a thickness of 500 $\mu$m. 
	For the lowest energy of 20 MeV at the Lintott spectrometer, electrons deposite 290 keV in a silicon strip, which corresponds to a charge of about 13 fC. For amplifying such small charges and for further processing, each 16 silicon strips are connected to a Gassiplex chip \cite{Santiard272783}. 
	Three Gassiplex chips are connected in a serial way, so that they act as a unit with 48 inputs and one output. 
	Two such 48-channel Gassiplex units are connected by a ribbon cable to one 96-channel silicon strip detector. 
	Because of the special configuration only 8 analog multiplexed outputs are needed to transmit channel level information from all 384 strips.
	Figure \ref{fig:lintott_detektor} displays a photograph of the system inside the vacuum chamber.	
	The Gassiplex chips require a special read-out scheme shown in \autoref{fig:gassiplex_readout}. 
	After a charge peaking time of 520 ns a Hold signal freezes all voltage levels at the inputs connected to the silicon strips. Next a Clock signal provided to the serializer produces a multiplexed signal with voltage levels at the output. Finally, after 2\ $\mu$s  needed for the baseline recovery, a Clear pulse sets the Gassiplex chip back into the operational mode.  

	\subsection{Previous Data Acquisition}
	
	In the previous data acquisition setup, the 8 cables with the analog multiplexed signals were connected to a special FPGA-based module described in Ref.~\cite{Lenhardt2006}. 
	It processes all the steps from the digitization of the multiplexed signal to the ready-to-use histogram containing detected electron positions.
	The communication with the module was realized via Telnet.
	The module also produced control signals for the Gassiplex read-out logic, triggered by the scintillator detector. 
	The complete setup is shown schematically in the left part of \autoref{fig:daq}.

	\section{New read-out process}

	\begin{figure*}
		\centering
		\includegraphics[width=\textwidth]{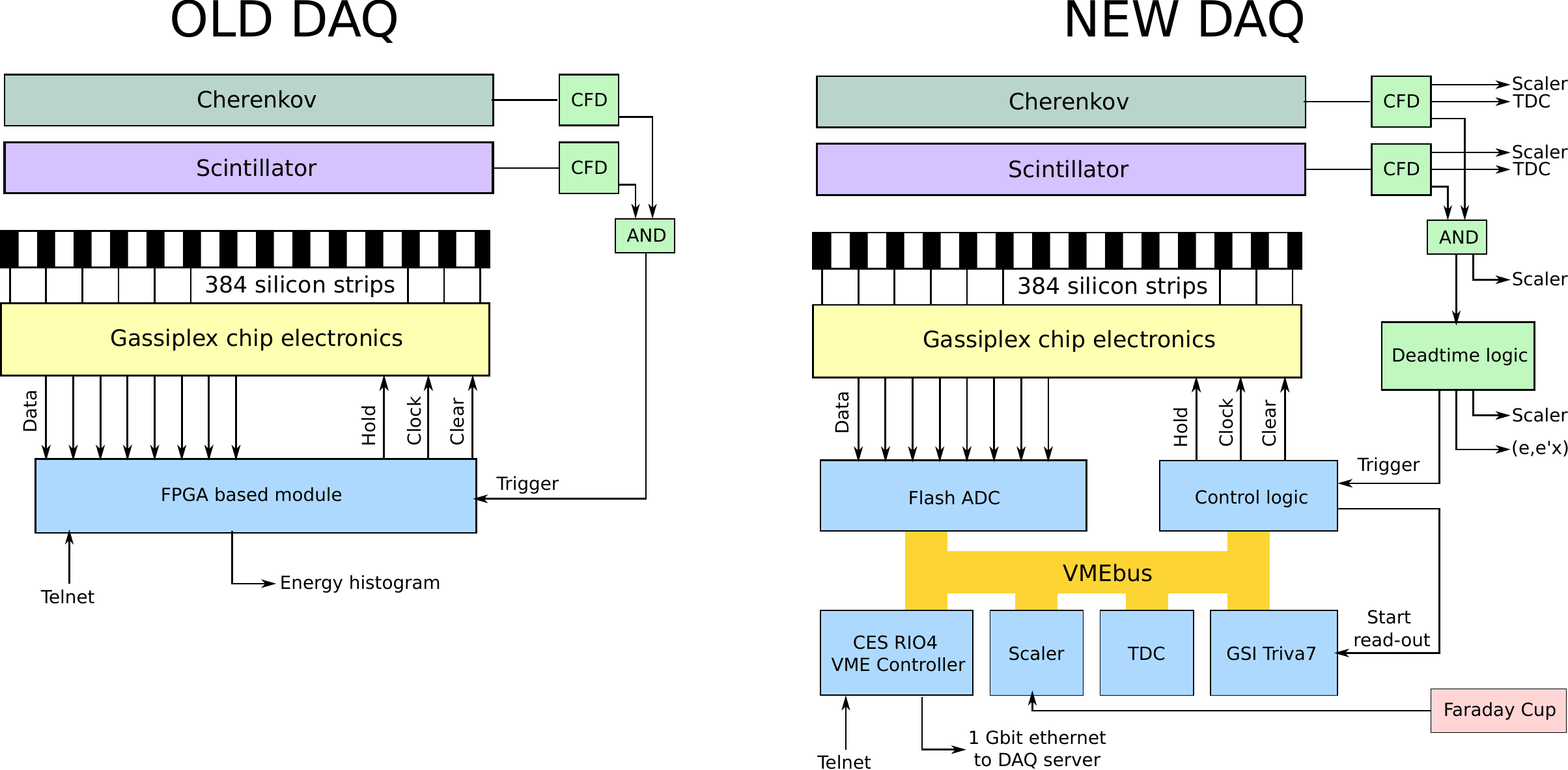}
		\caption{Scheme of the old (left) and new (right) data acquisition. 
			The trigger logic of the new system is realized with NIM logic modules. 
			Data are extracted with VME modules.
			For details see text.}
		\label{fig:daq}
	\end{figure*}
	
	The data acquisition setup described in this work is schematically shown in the r.h.s.\ of Fig.~\ref{fig:daq}. It starts with the analog multiplex signal coming from the Gassiplex chips, hence it replaces the FPGA-based module.
	The trigger coincidence logic and the dead-time logic are realized with NIM modules. 
	To create the necessary control signals for the Gassiplex electronics a VME module Wiener MDGG8 \cite{WienerMDGG8} was used. 
	It provides a number of built-in logic gates that can be connected to inputs and outputs by setting corresponding VME registers. 
	Access to the configuration registers and the read-out process of the VME ADC and TDC modules is realized with the Multi Branch System (MBS) \cite{GSIMBS}. 
	The data is written to a file server and analyzed by a custom software.
	
	\subsection{Gassiplex control process}
	
	The control scheme for the Gassiplex chips shown in \autoref{fig:gassiplex_readout} is realized with a multi-purpose Wiener MDGG8 module. 
	First, a coincidence signal between the scintillator and the Cherenkov detector is passed through the dead-time logic to an input of the MDGG8 module.  
	After a peaking time of 520 ns required by the Gassiplex chips, the MDGG8 module produces a Hold signal to freeze all voltage levels of the silicon strips. 
	At the same time a Clock signal consisting of 48 pulses with a period length of 144 ns is generated to produce a multiplexed signal of 48 channel levels at all 8 analog outputs.
	Finally, after a 2\ $\mu$s long baseline recovery pause a 80 ns long Clear pulse sets the Gassiplex chip back to live mode.

	\subsection{Data read-out}
	
	The differential multiplexed signals from the Gassiplex electronics are connected to a 16-bit Flash-ADC VME module Struck SIS3302 \cite{StruckSIS3302} with 100 $\Omega$ differential inputs. 
	The internal clock frequency of the ADC was set to 25 MHz to reduce the number of sampling points. 
	The number of samples to be digitized was set to 184 corresponding to a trace length of 7360 ns and 368 Byte per channel and trigger.
	
	The trigger signal starts the read-out process by activating the GSI TRIVA7 module \cite{GSITRIVA7} whose status is polled by MBS running on a CES RIO4-8072 VME master controller \cite{RIO4}. 
	First, the scaler and TDC data are read and then the digitized traces. 
	The Flash-ADC provides the possibility to write and read the data to two independent memory pages for every input channel. 
	A memory page is 8 MByte deep sufficient to store $>$20k events. 
	With every read-out event the memory page is switched and the data from the last active memory page can be read using multi-block-transfer mode. 
	The binary data stream from the VME modules is transported over an Ethernet connection to a local data acquisition server, where the data is written to so-called list mode data (LMD) files.
	
	\subsection{Live time}
	
	The live time of the system is defined by two factors. 
	The first one is the time needed by the Gassiplex electronics to serialize the analog channel levels to a multiplex signal, whose length is about 9.5\ $\mu$s including 144\ ns/ch $\times$ 48 ch for the desired analog signal, gaps after 16 and 32 channels and the baseline recovery time. 
	To cover this period of time a fixed, non-extendable dead time of 10\ $\mu$s length was introduced by a back-looped logic directly after the trigger detector coincidence unit.
	
\begin{figure}
		\centering
		\includegraphics[width=0.45\textwidth]{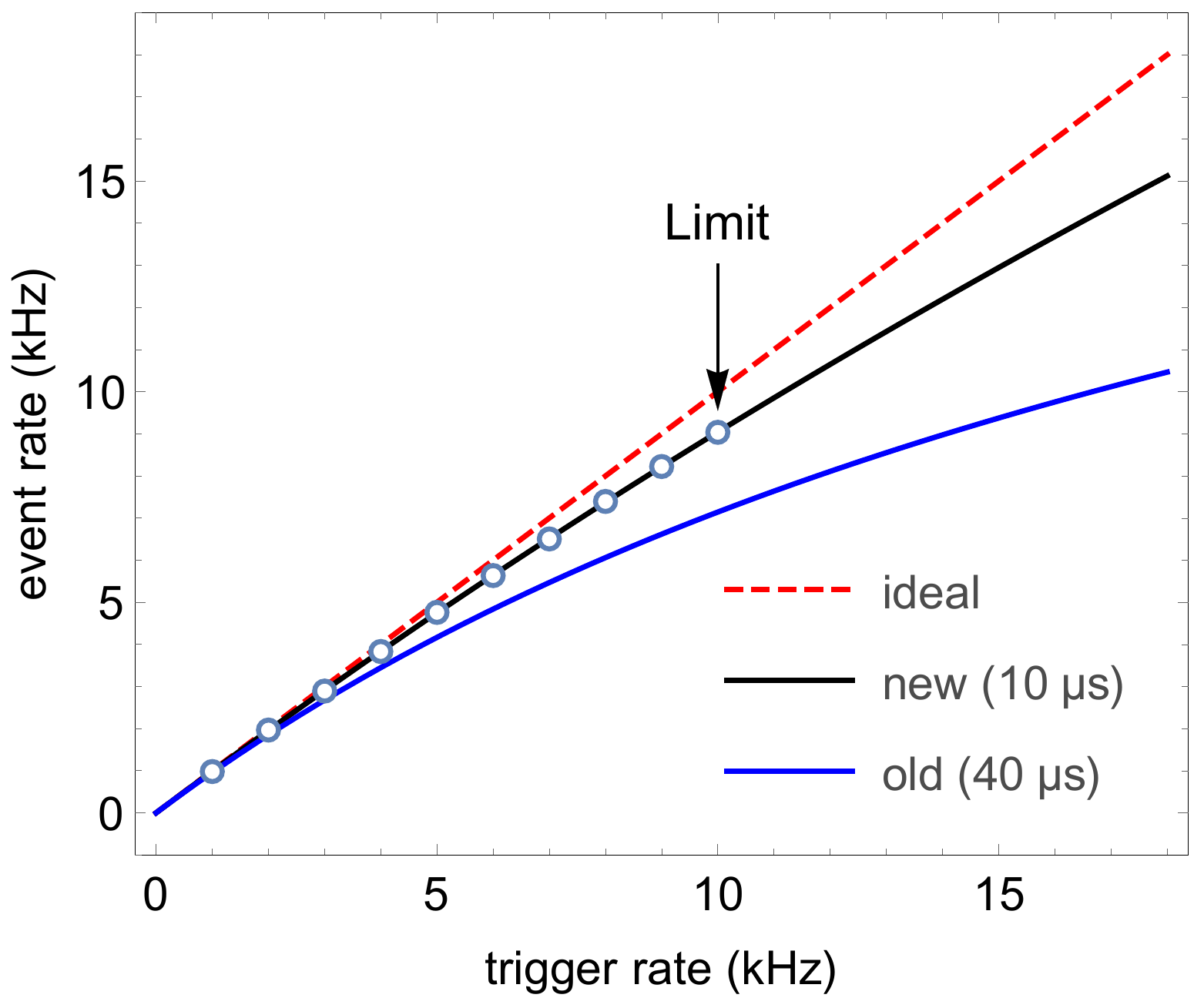}
		\caption{Measured event rate for the new data acquisition. The maximum event rate is limited by the read-out data rate.}
		\label{fig:eventrate_measured}
	\end{figure}

	The second factor is the transfer rate of the VME bus between the digitizer and the VME controller measured to be about 30 MByte/s. 
	It introduces a limit for the maximum event rate as illustrated in \autoref{fig:eventrate_measured} showing the event rate as a function of the trigger rate for the old and the new system.  
	%The value of the total live time is given by the fraction of the scaler value from the coincidence unit and the number of digitized events
	For a dead time fraction of 10\%,  the dead time of 10 $\mu$s corresponds to a maximum event rate of about 10 kHz, which is more than sufficient for most electron-scattering experiments at the S-DALINAC.
	
	\subsection{Beam current measurement}
	
	After passing the scattering target the electron beam is collected in a Faraday cup to measure the beam current. 
	The Faraday cup is connected via an BNC cable to an in-house made linear current-to-frequency converter that produces square pulses with a frequency depending on the measured beam current. 
	The conversion factor, selected by a resistor on the converter PCB, was chosen to 101.6\ $\mu$A\ /\ 32.768\ MHz to cover a range from 0.1 nA to 10 $\mu$A. 
	The output of this converter is connected to a VME scaler that counts the number of pulses since the last read-out, which corresponds to the collected charge.
	Different to the old data acquisition system the measurement of the beam current was integrated into the new data acquisition to have all relevant values in one data file. 
	
	\subsection{Expandability for (e,e$^\prime$x) Coincidence Experiments}
	
	To achieve the best signal-to-noise ratio and hence reduce the beam time in a coincidence (e,e$^\prime$x) experiment it is most important to optimize the resolution of time coincidence peak between the detected electron and the coincidence detectors.  
	A TDC module with internal ring buffer and 30 ps (RMS) time resolution was implemented into the current setup. 
	The purpose of this module is to measure the arriving times of the trigger detector signals with respect to the trigger signal after the dead-time logic, which serves as a reference signal and is provided to the read-out system of coincidence detectors. 
	By comparing the electron arriving times and the coincidence detector times to the reference signal a coincidence histogram can be built. 
	
	The current trigger detector setup consists of a 0.5 cm thin plastic scintillator detector and a Cherenkov detector with a thickness of 5 cm. Each detector is read out by one photo-multiplier. Because of this configuration the time resolution of the trigger signal is limited by the size of the scintillator to $\approx 8$ ns. To overcome this limitation a new setup for the trigger detector is presently under development. It will consists of two 1 cm thick scintillators, each with two read-out detectors that will improve the time resolution to about $1.3$ ns (FWHM).
	The properties of the TDC module take this development into account.  
	
	\section{Data processing}
	
	\subsection{Data extraction}
	
	The binary data from the VME master controller is transferred over an Ethernet connection to a data acquisition server, where it is written to LMD files. 
	For further processing an application written in C++ with a graphical user interface realized with Qt \cite{qt}. It covers all necessary steps from the binary LMD files to the summarized experimental data presented to an experimenter. 
	A graphical user interface allows to perform basic data analysis like fitting a model function to excited states in the energy spectrum.

	\begin{figure}[h]
		\centering
		\includegraphics[width=0.35\textwidth]{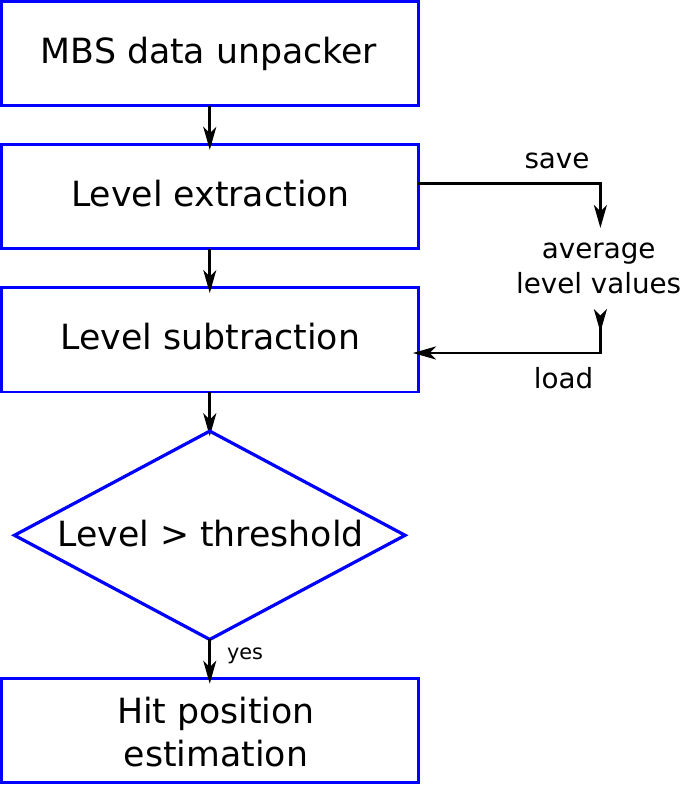}
		\caption{Simplified scheme of the processing of the data stream from the data acquisition.}
		\label{fig:dataprocessing}
	\end{figure}
\begin{figure}%[h]
		\centering
		\includegraphics[width=0.45\textwidth]{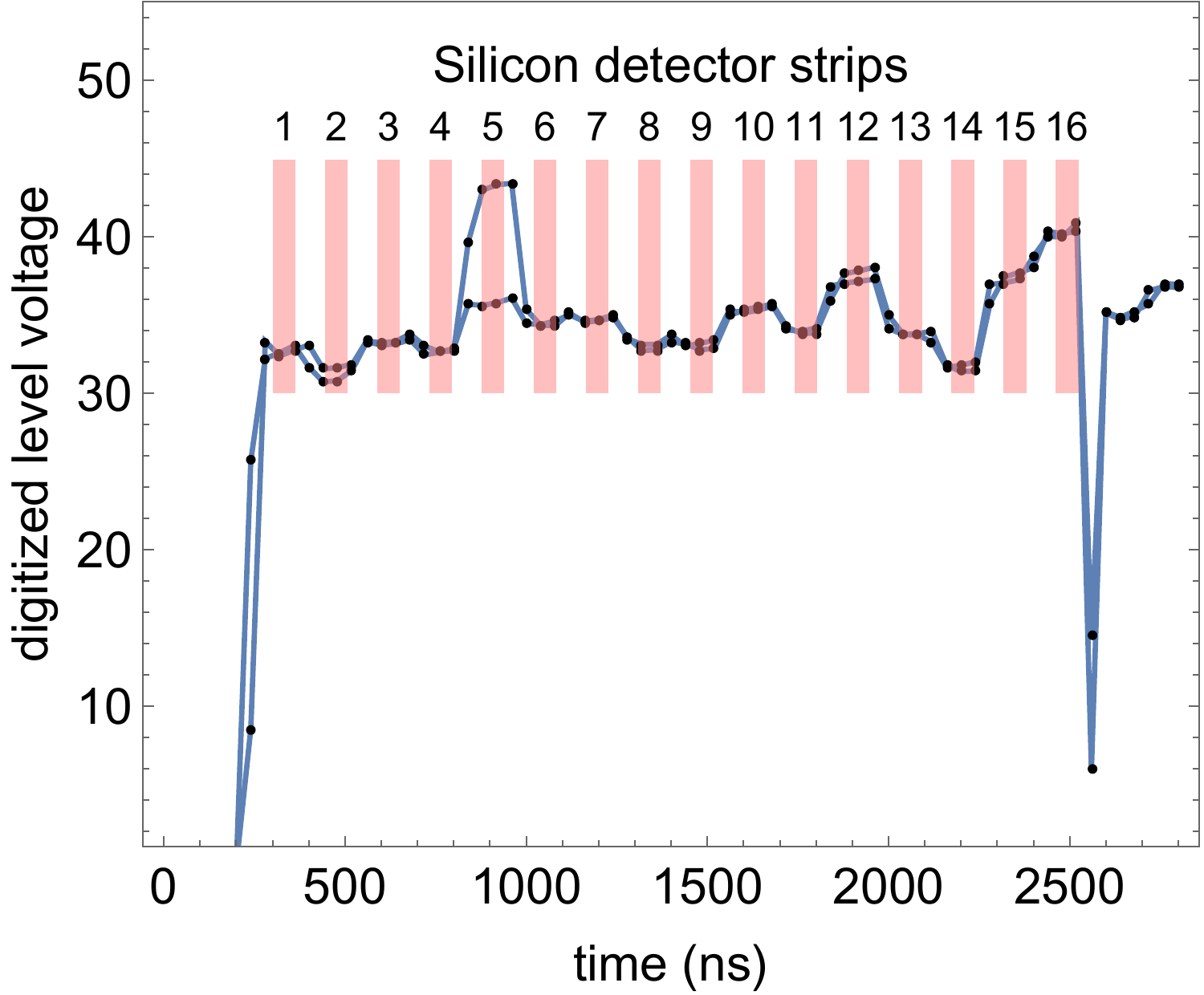}
		\caption{Example of two digitized multiplexed signals from the Gassiplex electronics without and with a hit (strip 5). The hit signal amplitude is about 250 mV in this case.
		Because of the low bandwidth the channel edges are fuzzy. 
	    To eliminate the uncertainties due to the edges only data points in the red marked areas are used for further processing.}
		\label{fig:multiplexedsignal}
	\end{figure}

	The most important steps in data processing built into the application are schematically shown in \autoref{fig:dataprocessing}. In the first step, the ADC data is extracted from MBS binary files to a data structure with eight lists of digitized multiplexed signals, representing the analog voltage levels of the silicon strips. Figure~\ref{fig:multiplexedsignal} presents an example of the digitized multiplexed signal from the Gassiplex electronics. 
	To extract the actual voltage levels, a software leading edge discriminator with a threshold set to half of the typical signal height was applied to find the starting point of the first level in the data. 
	
	In the next step, an iterative routine assigns sampling points to the actual silicon strips.
	In this procedure only the central part of a voltage level region with a length of 64 ns is used to avoid problems with rising and falling edges due to the limited bandwidth of the signal driver of the Gassiplex chip electronics.
	Considering the Gassiplex clock period length of 144 ns and the sampling point distance of 40 ns, there is always at least one sampling point for each silicon strip. When two sampling points are available, then the average value of both is assigned to a channel. 
	For each channel a histogram is produced to monitor the voltage levels. It mostly contains values without an electron signal and some increased voltage levels for situations when electrons have hit the region close to this particular channel of the detector. 
	After recording several thousand events, the  maximum value recorded is used as the default voltage level for that specific channel. A list of default voltage levels is written into a text file and can be used for upcoming experiments.
	
	\subsection{Estimation of the hit position}
	
\begin{figure}%[h]
		\centering
		\includegraphics[width=0.45\textwidth]{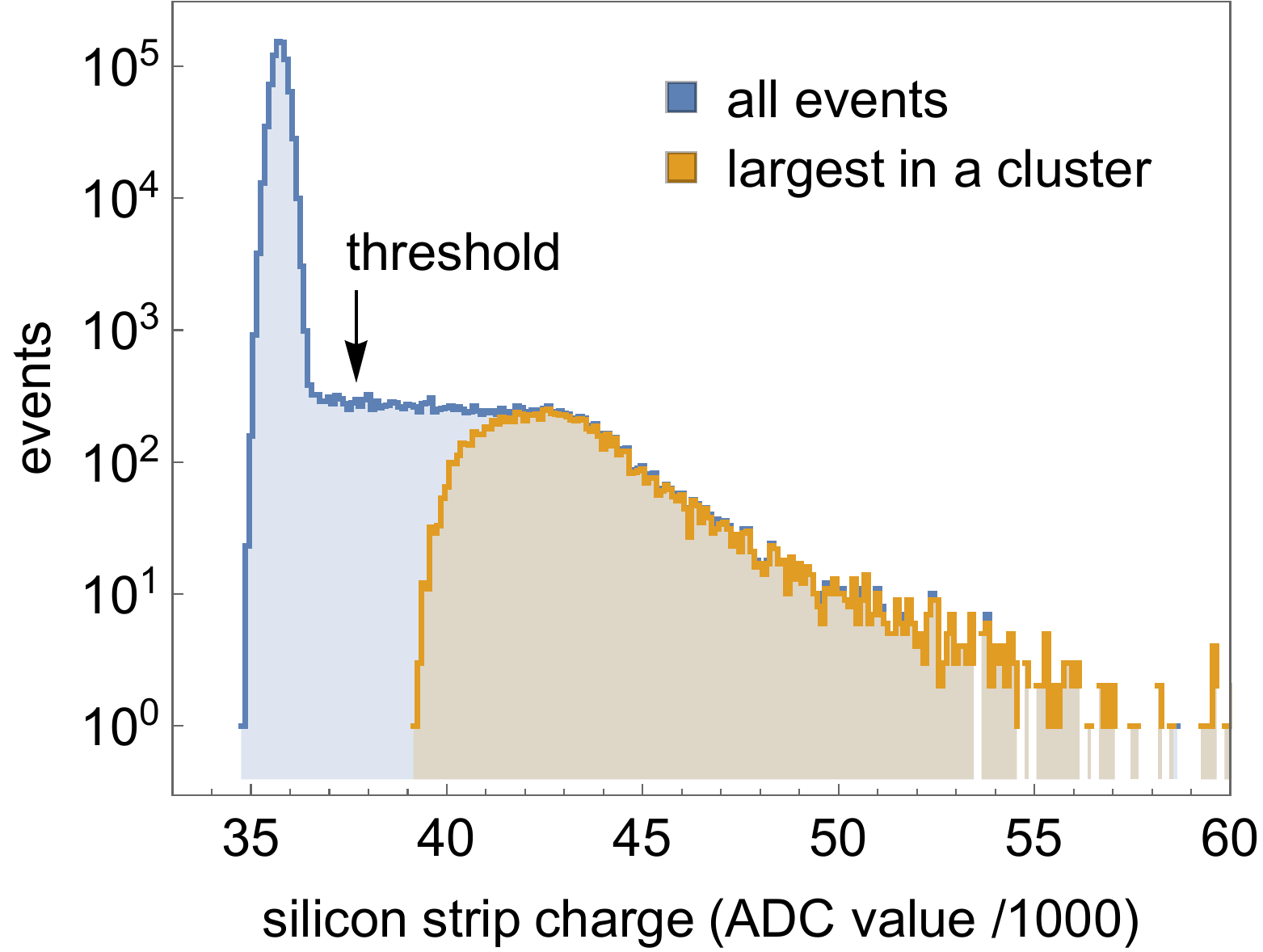}
		\caption{Blue: Histogram for ADC values of a single silicon strip. 
			The maximum of the Gaussian peak provides the default voltage level. 
			Because an electrons can deposit energy in two or three silicon strips, the charge measured for one strip is continuous, causing a flat plateau in the histogram. 
			Orange: Distribution for strips with the largest charge in a detected cluster.}
		\label{fig:silicon_strip_threshold}
	\end{figure}

	For the identification of an electron hit on the focal plane the default level values are subtracted from the measured ones (cf.\ \autoref{fig:dataprocessing}). 	Next, every channel is compared to a threshold set to ten times the standard deviation of the channel level distribution. 
	This value is high enough to avoid an erroneous triggering of the algorithm but low enough not to exclude any true events as demonstrated in \autoref{fig:silicon_strip_threshold}. 

	Because electrons impinge on the detector under an average angle of $34^\circ$, usually two or three silicon strips show a signal. 
	This provides additional information and allows to calculate a continuous value for the electron hit position, causing a flat plateau in \autoref{fig:dataprocessing}. 
	Its value is determined in the following way.
	First the algorithm is looking for a silicon strip with a signal above threshold. 
	For a true signal an additional cluster search in the neighboring $\pm2$ strips around is performed to find the strip with the largest value in the cluster ($\textmd{ch}_\textmd{l}$). 
	In the next step we assume that $\textmd{val}(\textmd{ch})$ -- the charge deposited in one silicon strip -- depends linearly  on the distance to that strip. 
	Then, the actual electron position can be calculated by linear interpolation normalized to the total charge of the two involved silicon strips
	\begin{equation}
	\textmd{ch}= \textmd{ch}_\textmd{l} + \textmd{p}
	\end{equation}
	with
	\begin{equation}
	\textmd{p} = \frac{-\textmd{val}(\textmd{ch}_\textmd{l}-1)}{\textmd{val}(\textmd{ch}_\textmd{l}-1) + \textmd{val}(\textmd{ch}_\textmd{l})}
	\end{equation}
	for $\textmd{val}(\textmd{ch}_\textmd{l}-1) >\textmd{val}(\textmd{ch}_\textmd{l}+1)$
	and 		
	\begin{equation}	
	\textmd{p} = 		\frac{\textmd{val}(\textmd{ch}_\textmd{l}+1)}{\textmd{val}(\textmd{ch}_\textmd{l}+1) + \textmd{val}(\textmd{ch}_\textmd{l})}
	\label{eq:hit_position}
	\end{equation}
	else.
	
	%	\begin{numcases}{\textmd{p} = }
	%		\frac{-\textmd{val}(\textmd{ch}_\textmd{l}-1)}{\textmd{val}(\textmd{ch}_\textmd{l}-1) + \textmd{val}(\textmd{ch}_\textmd{l})}, & 
	%		\text{if }
	%		$ \!\begin{aligned}[t]
	%		&\textmd{val}(\textmd{ch}_\textmd{l}-1)\\
	%		>&\textmd{val}(\textmd{ch}_\textmd{l}+1)
	%		\end{aligned}$
	%		\\
	%		\frac{\textmd{val}(\textmd{ch}_\textmd{l}+1)}{\textmd{val}(\textmd{ch}_\textmd{l}+1) + \textmd{val}(\textmd{ch}_\textmd{l})}, & else.
	%		\label{eq:hit_position}
	%	\end{numcases}
	
\begin{figure}[t]
		\centering		\includegraphics[width=0.45\textwidth]{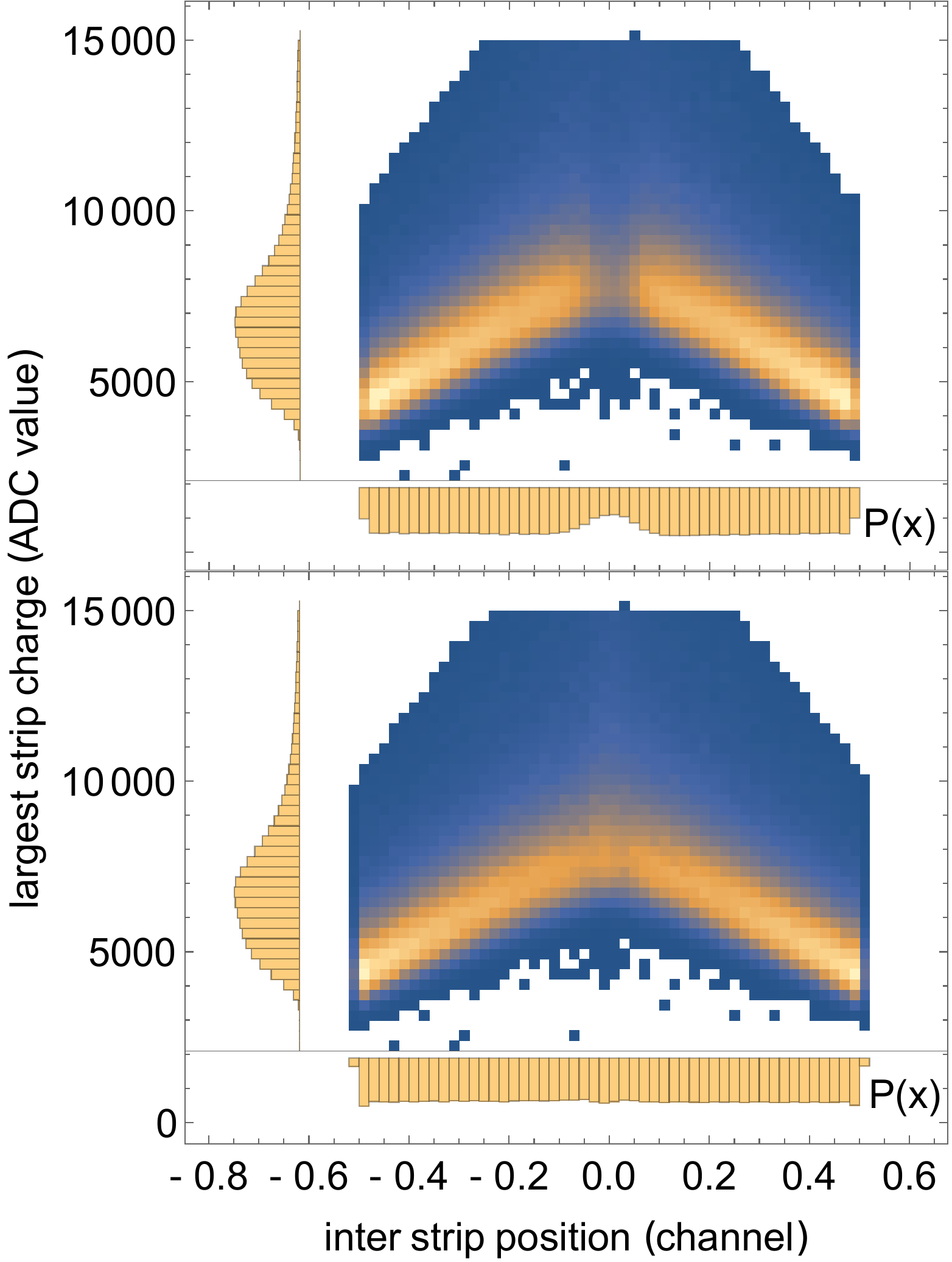}
		\caption{Density plot of the calculated continuous electron hit position around a discrete silicon strip on the x-axis before (top figure) and after the correction by \autoref{eq:position_correction_1} or (\ref{eq:position_correction_2}). The y axis shows the measured amplitude of the highest silicon strip inside a cluster. 
			The uncorrected case shows a dent around zero in the histogram subfigure that disappears after the correction. 
			The width of the distribution between $\pm 0.5$ and 0.0 corresponds to the achieved position resolution. }
		\label{fig:inter_strip_pos_vs_max_strip_charge_uncorrected}
	\end{figure}

	The upper part of \autoref{fig:inter_strip_pos_vs_max_strip_charge_uncorrected} shows the resulting density histogram for inter-strip positions $P(x)$ versus the highest strip amplitude inside a cluster. 
	The distribution $P(x)$ is expected to be flat but exhibits a dent around zero. 
	This is the region where most of the total charge is deposited in a single strip only. 	
	Nevertheless, the neighbor strips have non-zero values due to the statistical variations in the level voltages, shifting the estimated position away from zero. 
	To correct this problem an nonlinear correction function is applied
	\begin{equation}
	\mathrm{p}_\mathrm{corr}(p) = 
	\frac{1}{\sum_{x=0.0}^{0.5} P(x)} \sum_{x=0.0}^{p} P(x)
		\label{eq:position_correction_1}
	\end{equation}
	if the position is $> 0$ and
	\begin{equation}
	\mathrm{p}_\mathrm{corr}(p) = 	
	\frac{1}{\sum_{x=0.0}^{-0.5} P(x)} \sum_{x=0.0}^{p} P(-x) 
	\label{eq:position_correction_2}
	\end{equation}
	else.
	%	%
	%	\begin{numcases}{\mathrm{p}_\mathrm{corr}(p) = }
	%		\frac{1}{\sum_{x=0.0}^{0.5} P(x)} \sum_{x=0.0}^{p} P(x), &\text{if $pos > 0$} \\
	%		\frac{1}{\sum_{x=0.0}^{-0.5} P(x)} \sum_{x=0.0}^{p} P(-x), & else.
	%		\label{eq:position_correction}
	%	\end{numcases}	
	The result of the correction is visible in the bottom part of  \autoref{fig:inter_strip_pos_vs_max_strip_charge_uncorrected}. 
	The inter strip positions are now flatly distributed.
	By this technique the achieved position resolution can be improved from 0.65 $\mu$m defined by the size of the silicon strips to 0.36 $\mu$m around zero and 0.18 $\mu$m towards $P(x) = \pm 0.5$.

	The improved position resolution can be utilized to achieve a better energy resolution because the bin size of the position histogram can be reduced. 
	The shape of the elastic line and transitions to excited states in electron scattering is usually described by an analytical model with five fit parameters \cite{Hofmann2002}. 
	Depending on the kinematics of the experiment, the width of the lines may correspond to $3-4$ channels only,
	% as illustrated by the example shown in the upper part of \autoref{fig:fine_position},
	creating problems with over-fitting and uncertainty estimations in the past. 
	The interpolated data with more channels per line shown in the bottom part makes the model fit algorithm more robust.
	
	%
%	\begin{figure}%[h]
%		\centering
%		\includegraphics[width=0.45\textwidth]{figures/used/fine-position.pdf}
%		\caption{Example of an elastic (left) and $\frac{7}{2}^{+}$-state inelastic line (right) of the $^\text{197}$Au(e,e$^\prime$) reaction for the discrete position estimation (top) and the continues one (bottom). %The interpolated data with more data points makes the model fit algorithm more robust.
%		}
%		\label{fig:fine_position}
%	\end{figure}
	
	\subsection{Automated beam energy shift correction}
	
\begin{figure}%[h]
		\centering
		\includegraphics[width=0.45\textwidth]{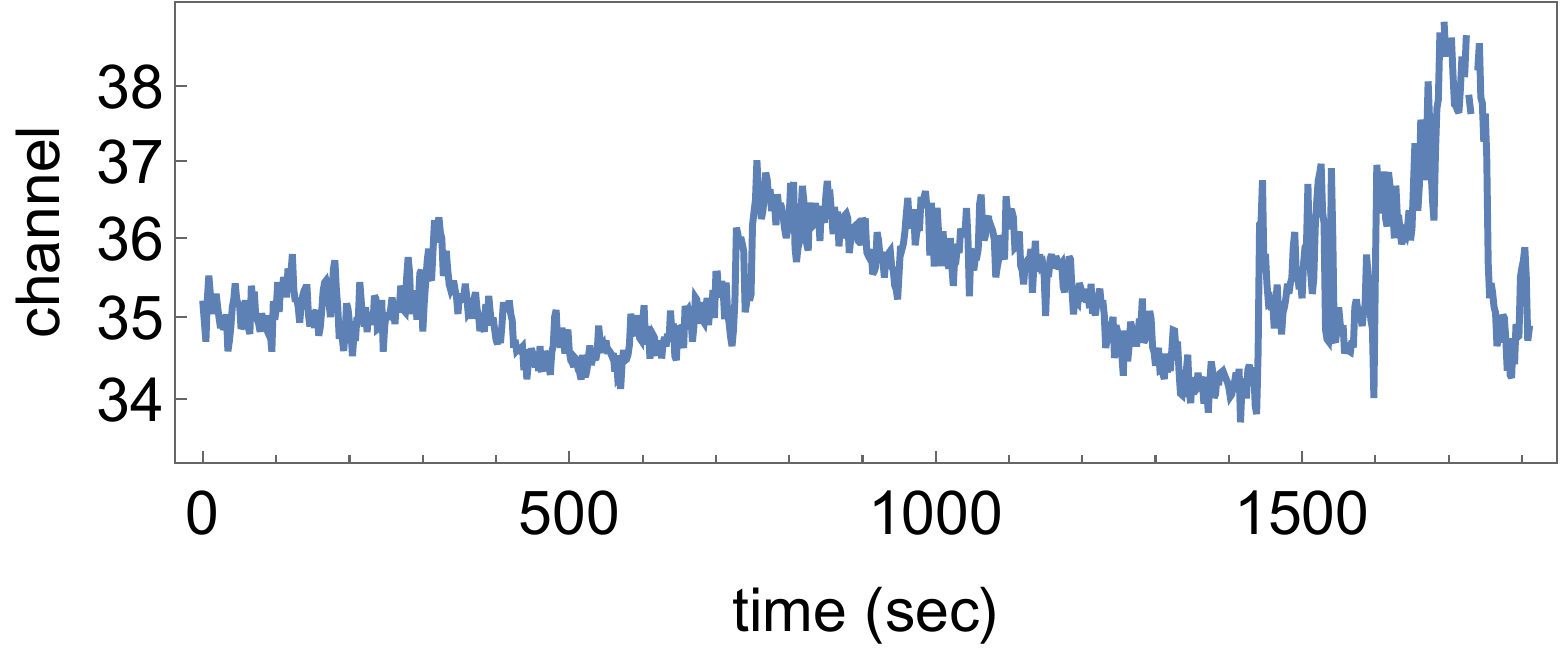}
		\caption{Estimated position of the elastic line in the focal plane detector during a commissioning run. 
			The time between two data points corresponds to approximately 0.5 seconds. 
			In the shown experiment one channel corresponds to 4.4 keV, so the position variations are of the order of the energy resolution.}
		\label{fig:position_over_time}
	\end{figure}

	The electron beam energies and thus the position of the elastic line in the detector plane shows fluctuations with time scales from minutes to sub-seconds. 
	The reasons for that are various, partially due to the complexity of the accelerator. Some important contributions are initial energy variations caused by the thermionic gun or even caused by long-time drifts due to thermal variations of the accelerator RF components. 
	Figure \ref{fig:position_over_time} demonstrates the phenomenon by a continuous measurement of the position of the elastic line during the commissioning experiment described below. 
	The channel position was estimated from histograms of the electron hit position using 200 events and taking the mean position of a Gaussian model function fitted to the data. 
	The number of 200 events was chosen because it is large enough for stable fit results and small enough to catch variations on time levels down to several hundreds of milliseconds (depending on the actual event rate).

	\begin{figure}[h]
		\centering
		\includegraphics[width=0.45\textwidth]{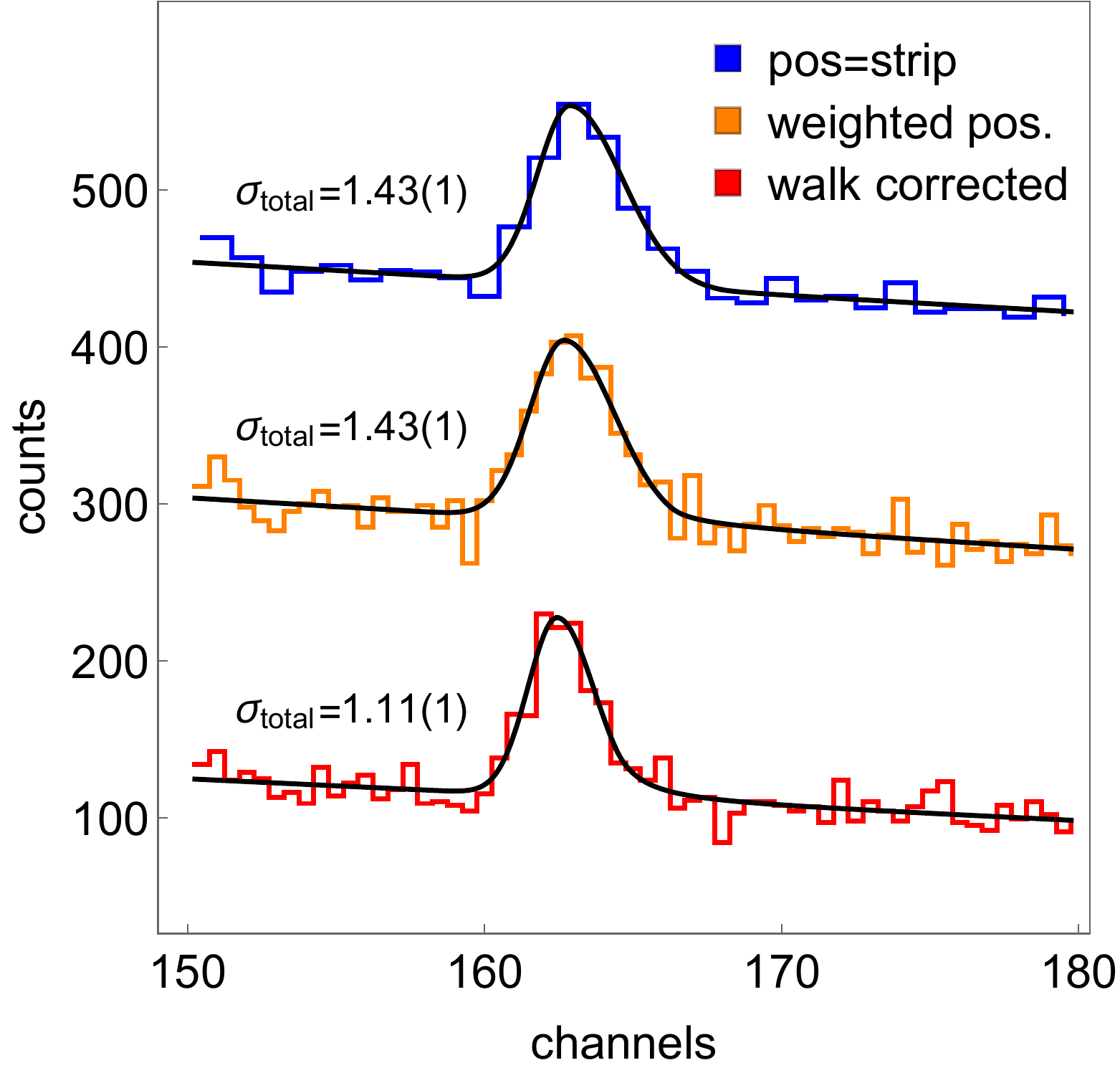}
		\caption{Exited $\frac{7}{2}^{+}$ state of the $^\text{197}$Au(e,e$^\prime$) reaction before (orange) and after the automated position walk correction (red) for the run shown in \autoref{fig:position_over_time}. The blue peak represents the old data acquisition for comparison.}
		\label{fig:walk_correction_and_weighted_hit_position_inelastic}
	\end{figure}
	
	With the new data acquisition an automated shift correction was introduced. 
	The position of the elastic line extracted from the  first histogram with 200 events is set as the initial position. 
	The position for the subsequent electron hits is shifted by the difference between the initial position and the mean position of the 200 event packages. 
	The benefit of this simple method is shown in \autoref{fig:walk_correction_and_weighted_hit_position_inelastic} for the first exited state in $^{197}$Au observed in the commissioning run. 
	The line width deduced from the corrected histogram (red) is improved by 22\ \% compared to the uncorrected spectrum (orange) and a spectrum corresponding to the old data acquisition (blue).
	%The exactly benefit is depending on the provided intrinsic electron beam stability in terms of energy and position variations on the target. 

	\section{Commissioning}
	
	\begin{figure}%[h]
		\centering
		\includegraphics[width=0.45\textwidth]{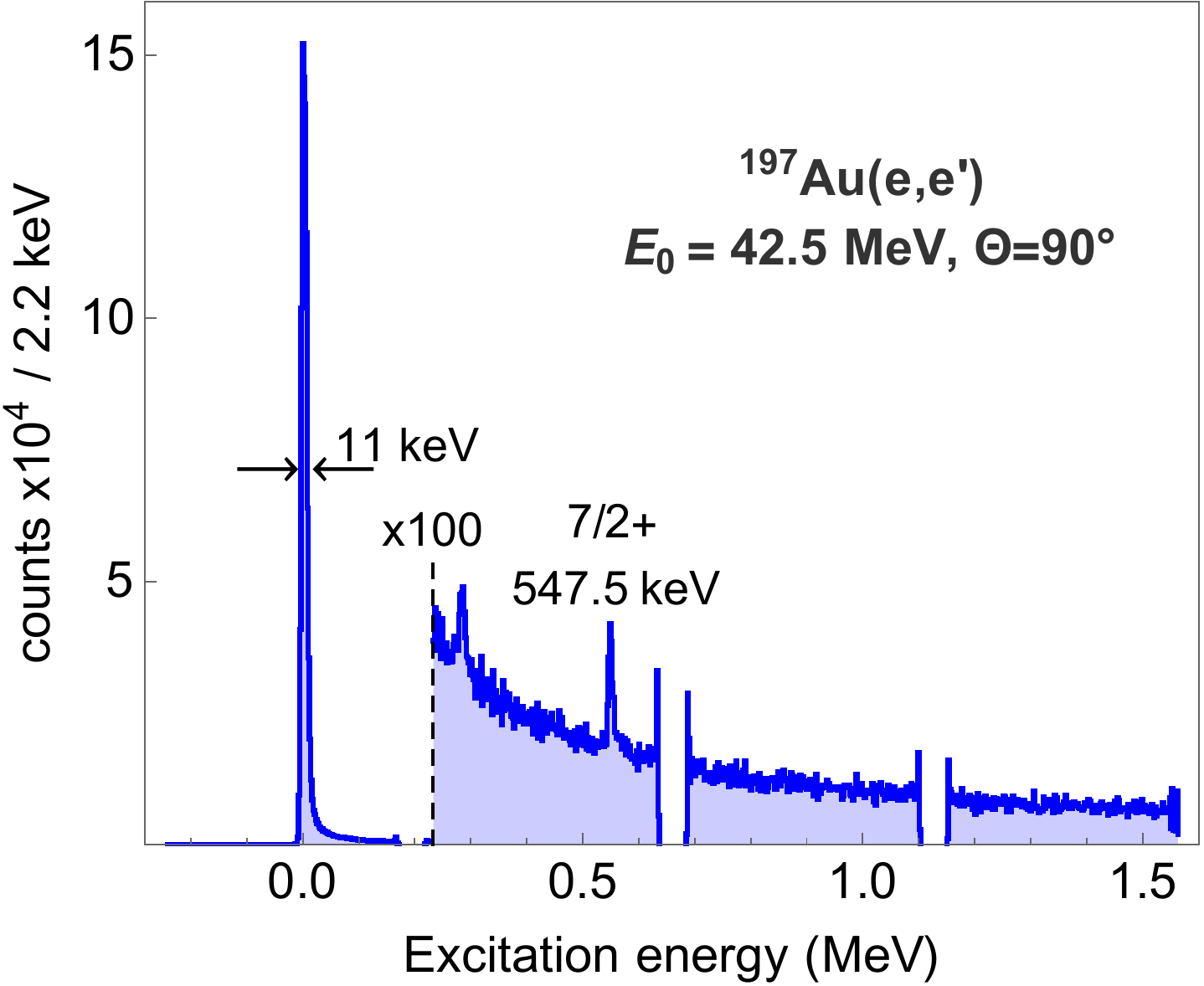}
		\caption{Excitation spectrum of the $^\text{197}$Au(e,e$^\prime$) reaction used for commissioning after an automated peak shift correction. 
			The gaps in the spectrum arise from the physical gaps between the four silicon strip detectorsvisible in \autoref{fig:lintott_detektor}.}
		\label{fig:spectrum}
	\end{figure}
	
	Commissioning of the new data acquisition was conducted with a $(42.5\pm0.5)$\ MeV electron beam impinging on a $^\text{197}$Au target with an areal density $\rho=1$\ mg/cm$^2$. 
	The Lintott spectrometer was placed at 90$^\circ$. 
	The measured excitation spectrum is shown in \autoref{fig:spectrum}. 
	Peaks corresponding to elastic scattering and due to the first excited $\frac{7}{2}^{+}$ state at 547.5\ keV in $^{197}$Au are visible. 
	The measured energy resolution of the elastic line was 11 keV (FWHM), a  value regularly achieved at the Lintott spectrometer since the installation of the new high energy scraper system \cite{Jurgensen2019}.
	
	\section{Summary}
	
	A new event-based data acquisition for the existing silicon strip focal plane detector at the Lintott magnetic spectrometer at the S-DALINAC was developed. 
	The system allows to conduct high resolution electron scattering experiments with event rates up to 10 kHz. Because of the event-based design and time information available from the trigger detectors, coincidence experiments are possible in future. 
	The event-based design provides also the possibility to detect and correct shifts in the energy spectrum, leading to significant improvements of the energy resolution. 
	
	\section*{Acknowledgments}
	
	We thank the accelerator crew at the S-DALINAC for providing excellent beams.
	This work was funded by the Deutsche Forschungsgemeinschaft
	(DFG, German Research Foundation) under Grant No.\ SFB 1245 (project ID 279384907) and GRK 2128 (project ID 264883531).

	
	
	
	\bibliographystyle{elsarticle-num}
	\bibliography{Lintott}

\begin{thebibliography}{10}
\expandafter\ifx\csname url\endcsname\relax
  \def\url#1{\texttt{#1}}\fi
\expandafter\ifx\csname urlprefix\endcsname\relax\def\urlprefix{URL }\fi
\expandafter\ifx\csname href\endcsname\relax
  \def\href#1#2{#2} \def\path#1{#1}\fi

\bibitem{Pietralla2018}
N.~Pietralla, {The Institute of Nuclear Physics at the TU Darmstadt}, Nucl.
  Phys. News 28(2) (2018) 4.

\bibitem{Heyde2010}
K.~Heyde, P.~von Neumann-Cosel, A.~Richter,
  \href{https://link.aps.org/doi/10.1103/RevModPhys.82.2365}{Magnetic dipole
  excitations in nuclei: Elementary modes of nucleonic motion}, Rev. Mod. Phys.
  82 (2010) 2365.

\bibitem{Schull1977}
D.~Sch\"ull, J.~Foh, H.-D. Gr\"af, H.~Miska, R.~Schneider, E.~Spamer,
  H.~Theissen, O.~Titze, T.~Walcher, {High resolution electron scattering
  facility at the Darmstadt linear accelerator (DALINAC). 3. Detector system
  and performance of the electron scattering apparatus}, Nucl. Instrum. Meth.
  153 (1978) 29.

\bibitem{Walcher1977}
T.~Walcher, R.~Frey, H.-D. Gr\"af, E.~Spamer, H.~Theissen, {High resolution
  electron scattering facility at the Darmstadt linear accelerator (DALINAC).
  2. Beam transport system and spectrometer (energy loss system)}, Nucl.
  Instrum. Meth. 153 (1978) 17.

\bibitem{Lenhardt2006}
A.~Lenhardt, U.~Bonnes, O.~Burda, P.~von Neumann-Cosel, M.~Platz, A.~Richter,
  S.~Watzlawik,
  \href{http://www.sciencedirect.com/science/article/pii/S0168900206004979}{{A
  silicon microstrip detector in a magnetic spectrometer for high-resolution
  electron scattering experiments at the S-DALINAC}}, Nucl. Instrum. Meth. A
  562 (2006) 320.

\bibitem{Chernykh2007}
M.~Chernykh, H.~Feldmeier, T.~Neff, P.~von Neumann-Cosel, A.~Richter,
  \href{https://link.aps.org/doi/10.1103/PhysRevLett.98.032501}{{Structure of
  the Hoyle state in $^{12}\mathrm{C}$}}, Phys. Rev. Lett. 98 (2007) 032501.

\bibitem{Chernykh2010}
M.~Chernykh, H.~Feldmeier, T.~Neff, P.~von Neumann-Cosel, A.~Richter,
  \href{https://link.aps.org/doi/10.1103/PhysRevLett.105.022501}{{Pair decay
  width of the Hoyle state and its role for stellar carbon production}}, Phys.
  Rev. Lett. 105 (2010) 022501.

\bibitem{Burda2010}
O.~Burda, P.~von Neumann-Cosel, A.~Richter, C.~Forss\'en, B.~A. Brown,
  \href{https://link.aps.org/doi/10.1103/PhysRevC.82.015808}{Resonance
  parameters of the first $1/{2}^{+}$ state in $^{9}\mathrm{Be}$ and
  astrophysical implications}, Phys. Rev. C 82 (2010) 015808.

\bibitem{Burda2007}
O.~Burda, N.~Botha, J.~Carter, R.~W. Fearick, S.~V. F\"ortsch, C.~Fransen,
  H.~Fujita, J.~D. Holt, M.~Kuhar, A.~Lenhardt, P.~von Neumann-Cosel,
  R.~Neveling, N.~Pietralla, V.~Y. Ponomarev, A.~Richter, O.~Scholten,
  E.~Sideras-Haddad, F.~D. Smit, J.~Wambach,
  \href{https://link.aps.org/doi/10.1103/PhysRevLett.99.092503}{High
  energy-resolution inelastic electron and proton scattering and the
  multiphonon nature of mixed-symmetry ${2}^{+}$ states in $^{94}\mathrm{Mo}$},
  Phys. Rev. Lett. 99 (2007) 092503.

\bibitem{Walz2011}
C.~Walz, H.~Fujita, A.~Krugmann, P.~von Neumann-Cosel, N.~Pietralla, V.~Y.
  Ponomarev, A.~Scheikh-Obeid, J.~Wambach,
  \href{https://link.aps.org/doi/10.1103/PhysRevLett.106.062501}{Origin of
  low-energy quadrupole collectivity in vibrational nuclei}, Phys. Rev. Lett.
  106 (2011) 062501.

\bibitem{Scheikh2013}
A.~Scheikh~Obeid, O.~Burda, M.~Chernykh, A.~Krugmann, P.~von Neumann-Cosel,
  N.~Pietralla, I.~Poltoratska, V.~Y. Ponomarev, C.~Walz,
  \href{https://link.aps.org/doi/10.1103/PhysRevC.87.014337}{{$E2$ strengths
  and transition radii difference of one-phonon ${2}^{+}$ states of
  ${}^{92}\mathrm{Zr}$ from electron scattering at low momentum transfer}},
  Phys. Rev. C 87 (2013) 014337.

\bibitem{Scheickh2014}
A.~Scheikh~Obeid, S.~Aslanidou, J.~Birkhan, A.~Krugmann, P.~von Neumann-Cosel,
  N.~Pietralla, I.~Poltoratska, V.~Ponomarev,
  \href{https://link.aps.org/doi/10.1103/PhysRevC.89.037301}{{$B(E2)$ strength
  ratio of one-phonon ${2}^{+}$ states of ${}^{94}\mathrm{Zr}$ from electron
  scattering at low momentum transfer}}, Phys. Rev. C 89 (2014) 037301.

\bibitem{Kremer2016}
C.~Kremer, S.~Aslanidou, S.~Bassauer, M.~Hilcker, A.~Krugmann, P.~von
  Neumann-Cosel, T.~Otsuka, N.~Pietralla, V.~Ponomarev, N.~Shimizu, M.~Singer,
  G.~Steinhilber, T.~Togashi, Y.~Tsunoda, V.~Werner, M.~Zweidinger,
  \href{https://link.aps.org/doi/10.1103/PhysRevLett.117.172503}{First
  measurement of collectivity of coexisting shapes based on type ii shell
  evolution: The case of $^{96}\mathrm{Zr}$}, Phys. Rev. Lett. 117 (2016)
  172503.

\bibitem{DAlessio2020}
A.~D'Alessio, T.~Mongelli, M.~Arnold, S.~Bassauer, J.~Birkhan, I.~Brandherm,
  M.~Hilcker, T.~H\"uther, J.~Isaak, L.~J\"urgensen, T.~Klaus, M.~Mathy, P.~von
  Neumann-Cosel, N.~Pietralla, V.~Ponomarev, P.~Ries, R.~Roth, M.~Singer,
  G.~Steinhilber, K.~Vobig, V.~Werner,
  \href{https://link.aps.org/doi/10.1103/PhysRevC.102.011302}{{Precision
  measurement of the $E2$ transition strength to the ${2}_{1}^{+}$ state of
  $^{12}\mathrm{C}$}}, Phys. Rev. C 102 (2020) 011302.

\bibitem{Jurgensen2019}
L.~Jürgensen, et~al., {High-Energy Scraper System for the S-DALINAC Extraction
  Beam Line - Commissioning Run}, in: {7th International Beam Instrumentation
  Conference}, 2019, p. MOPB03.

\bibitem{Santiard272783}
J.-C. Santiard, W.~Beusch, S.~Buytaert, C.~C. Enz, E.~H.~M. Heijne, P.~Jarron,
  F.~Krummenacher, K.~Marent, F.~Piuz,
  \href{https://cds.cern.ch/record/272783}{{Gassiplex: a low-noise analog
  signal processor for readout of gaseous detectors}}, Tech. Rep.
  CERN-ECP-94-17, CERN, Geneva (1994).
\newline\urlprefix\url{https://cds.cern.ch/record/272783}

\bibitem{WienerMDGG8}
\href{https://www.wiener-d.com/sc/modules/vme--modules/mdgg-8.html}{Wiener
  MDGG8 VME Module Product Page}.
\newline\urlprefix\url{https://www.wiener-d.com/sc/modules/vme--modules/mdgg-8.html}

\bibitem{GSIMBS}
\href{https://www.gsi.de/en/work/research/experiment_electronics/data_processing/data_acquisition/mbs.htm}{GSI
  Multi Brunch System Product Page}.
\newline\urlprefix\url{https://www.gsi.de/en/work/research/experiment_electronics/data_processing/data_acquisition/mbs.htm}

\bibitem{StruckSIS3302}
\href{https://www.struck.de/sis3302.htm}{Struck SIS3302 VME Module Product
  Page}.
\newline\urlprefix\url{https://www.struck.de/sis3302.htm}

\bibitem{GSITRIVA7}
\href{https://www.gsi.de/en/work/research/experiment_electronics/digital_electronic/digital_electronics/modules/vme/triva/triva7.htm}{GSI
  TRIVA7 VME Module Product Page}.
\newline\urlprefix\url{https://www.gsi.de/en/work/research/experiment_electronics/digital_electronic/digital_electronics/modules/vme/triva/triva7.htm}

\bibitem{RIO4}
\href{https://www.mrcy.com/siteassets/product-datasheets/mission-computing-avionics/single-board-computer-rio4-8072-datasheet.pdf}{Mercury
  Systems RIO4-8072}.
\newline\urlprefix\url{https://www.mrcy.com/siteassets/product-datasheets/mission-computing-avionics/single-board-computer-rio4-8072-datasheet.pdf}

\bibitem{qt}
\href{https://www.qt.io/}{Qt5 Framework Product Page}.
\newline\urlprefix\url{https://www.qt.io/}

\bibitem{Hofmann2002}
F.~Hofmann, P.~von Neumann-Cosel, F.~Neumeyer, C.~Rangacharyulu, B.~Reitz,
  A.~Richter, G.~Schrieder, D.~I. Sober, L.~W. Fagg, B.~A. Brown, {Magnetic
  dipole transitions in $^{32}$S from electron scattering at 180$^\circ$},
  Phys. Rev. C 65 (2002) 024311.

\end{thebibliography}
	
	
	
	
	%%\bibitem{}
	
	%%\end{thebibliography}
\end{document}